%% file: 11d-arxiv.tex
\documentclass[12pt]{article}
\usepackage[body={18cm, 23cm, centered}]{geometry}
\usepackage{
	amsmath,
	amssymb,
	amsthm,
	amscd,
	cite, 
	amsfonts,
	indentfirst,
	color,
	setspace,
	bbold,
	MnSymbol
}

\usepackage[debug,pageanchor=false]{hyperref}
\hypersetup{colorlinks=true,linktocpage,breaklinks,
	urlcolor=blue,
	linkcolor=red,
	citecolor=blue
}

\usepackage{tikz}
\usetikzlibrary{arrows}
\usetikzlibrary{shapes.misc}

\tikzset{cross/.style={cross out, draw=black, minimum size=2*(#1-\pgflinewidth), inner sep=0pt, outer sep=0pt},
	cross/.default={5pt}}
\usepackage[ normalem]{ulem}
\usepackage{tocloft}

\numberwithin{equation}{section}

\selectfont

\input{Defs.tex}

\begin{document}
	\renewcommand{\refname}{\begin{center}References\end{center}}
	
	\begin{titlepage}
		
		\vfill
		\begin{flushright}
			APCTP Pre2019-015\\
			IPM/P-2019/022
		\end{flushright}
		
		\vfill
		
		\begin{center}
			\baselineskip=16pt
			{\Large \bf 
				Tri-vector deformations in $d=11$ supergravity
			}
			\vskip 1cm
			Ilya Bakhmatov$^{a,d}$\footnote{\tt ilya.bakhmatov@apctp.org},
			Nihat Sadik Deger$^b$\footnote{\tt sadik.deger@boun.edu.tr},
			Edvard T. Musaev$^{c,d}$\footnote{\tt musaev.et@phystech.edu},\\
			Eoin \'O Colg\'ain$^{a,e}$\footnote{\tt ocolgain.eoin@apctp.org}, 
			Mohammad M. Sheikh-Jabbari$^{f,g}$\footnote{\tt jabbari@theory.ipm.ac.ir}
			\vskip .3cm
			\begin{small}
				{\it 
					$^a$Asia Pacific Center for Theoretical Physics, Postech, Pohang 37673, Korea\\
					$^b$Department of Mathematics, Bo\u{g}azi\c{c}i University, 34342, Bebek, Istanbul, Turkey\\
					$^c$Moscow Institute of Physics and Technology,
					Institutskii per. 9, Dolgoprudny, 141700, Russia\\
					$^d$Kazan Federal University, Institute of Physics, Kremlevskaya 16a, Kazan, 420111, Russia\\
					$^e$Department of Physics, Postech, Pohang 37673, Korea\\
					$^f$School of Physics, Institute for Research in Fundamental Sciences (IPM),\\ P.O.Box 19395-5531, Tehran, Iran\\
					$^g$ The Abdus Salam ICTP, Strada Costiera 11, 34151 Trieste, Italy
				}
			\end{small}
		\end{center}
		
		\vfill 
		\begin{center} 
			\textbf{Abstract}
		\end{center} 
		\begin{quote}
			We construct a $d=11$ supergravity analogue of the open-closed string map in the context of SL(5) Exceptional Field Theory (ExFT). The deformation parameter tri-vector $\Omega$ generalizes the non-commutativity bi-vector parameter $\Theta$ of the open string. When applied to solutions in $d=11$, this map provides an economical way of performing TsT deformations, and may be used to recover $d=10$ Yang-Baxter deformations after dimensional reduction. We present a generalization of the Classical Yang-Baxter Equation (CYBE) for rank 3 objects, which emerges from $d=11$ supergravity and the SL(5) ExFT. This equation is shown to reduce to the $d=10$ CYBE upon dimensional reduction. 
		\end{quote} 
		\vfill
		\setcounter{footnote}{0}
	\end{titlepage}
	
	\clearpage
	\setcounter{page}{2}
	
	\tableofcontents

	\section{Introduction}
We have witnessed swift progress in our understanding of integrable deformations of string $\sigma$-models over recent years, driven in no small part by the connection to dualities manifest at the level of the string $\sigma$-model. These ``dualities" include bona fide symmetries, such as Abelian T-duality \cite{Buscher:1987sk, Buscher:1987qj}, but have also steadily grown to encompass esoteric counterparts, notably non-Abelian  \cite{delaOssa:1992vc} and Poisson-Lie T-duality \cite{Klimcik:1995dy}, which are only symmetries at the one-loop level. Nevertheless, it is truly remarkable that $\eta$-deformations \cite{Klimcik:2002zj, Klimcik:2008eq} -- more generally Yang-Baxter (YB) deformations \cite{Delduc:2013fga, Delduc:2013qra, Kawaguchi:2014qwa} --  and $\lambda$-deformations \cite{Sfetsos:2013wia, Hollowood:2014qma} can all be absorbed into Poisson-Lie symmetric $\sigma$-models, called $\mathcal{E}$-models \cite{Klimcik:2015gba} (see also \cite{Hoare:2015gda, Sfetsos:2015nya}). 

In addition to various guises related via T-duality, the $\sigma$-model possesses another equivalent description as a target space-time geometry or supergravity solution. In this parallel setting, integrable deformations are $O(d,d)$ transformations.\footnote{Non-Abelian T-duality can be expressed in terms of $O(d,d)$ \cite{Hassler:2017yza,Lust:2018jsx, Sakatani:2019jgu,Catal-Ozer:2019hxw,Catal-Ozer:2019tmm}, thus bringing related developments in this direction under a common umbrella \cite{Hoare:2016wsk,Borsato:2016pas, Borsato:2017qsx, Borsato:2018idb}.} To the extent of our knowledge, this connection was initially observed in \cite{CatalOzer:2005mr} when multi-parameter Lunin-Maldacena (TsT) transformations \cite{Lunin:2005jy,Frolov:2005dj}, the simplest examples in the YB class, were recast as $O(d,d)$ transformations with \textit{constant} entries. In this reformulation, the $O(d,d)$ is specified by a (constant) bi-vector $\Theta$, but even in the absence of integrability, one has the freedom to replace the bi-vector by an $r$-matrix solution to the Classical Yang-Baxter Equation (CYBE) \cite{Araujo:2017jkb, Araujo:2017jap, Araujo:2017enj} and this guarantees the deformation is still a solution of supergravity \cite{Bakhmatov:2018apn}, more accurately generalized supergravity \cite{Arutyunov:2015mqj, Wulff:2016tju}.\footnote{See \cite{Hong:2018tlp} for an account of how these equations originally appeared in anomalous non-Abelian T-duality transformations \cite{Elitzur:1994ri}.} Curiously, the $O(d,d)$ transformation is nothing other than the open-closed string map of Seiberg \& Witten \cite{Seiberg:1999vs} from noncommutativity in string theory.  

Once reformulated in terms of $O(d,d)$, YB deformations were quickly embedded \cite{Sakamoto:2017cpu, Fernandez-Melgarejo:2017oyu, Sakamoto:2018krs} in Double Field Theory (DFT) \cite{Siegel:1993th, Siegel:1993xq, Hull:2009mi,Hohm:2010pp}, a setting where this symmetry is manifest. Moreover, the equations of motion of generalized supergravity were also recovered from DFT \cite{Sakatani:2016fvh} and Exceptional Field Theory (ExFT) \cite{Baguet:2016prz}. Moving beyond toroidal backgrounds, DFT was extended to encompass group manifolds \cite{Blumenhagen:2014gva, Blumenhagen:2015zma} with manifest Poisson-Lie T-duality symmetry \cite{Hassler:2017yza}. Ultimately, these developments paved the way for Poisson-Lie T-dualisable $\sigma$-models on curved spaces \cite{Demulder:2018lmj}, thus completing a remarkable circle back to the $\mathcal{E}$-models.  

Throughout these developments one thing has been missing, namely a method to perform deformations directly in M-theory, or $d=11$ supergravity. Here, we begin to address this concern. In essence, one is looking for a generalization of the open-closed string map, but with the NSNS two-form $B$ and corresponding bi-vector, $\Theta$ or equivalently $\beta$ in the notation of \cite{Andriot:2013xca, Bakhmatov:2018bvp}, replaced with the three-form potential $C$ and tri-vector $\Omega$. Despite difficulties defining noncommutativity for membranes \cite{Bergshoeff:2000jn, Kawamoto:2000zt, Bergshoeff:2000ai}, it turns out that the required map or transformation already exists in the literature \cite{Blair:2014zba} as a frame change in ExFT between geometric and non-geometric frames (see \cite{Andriot:2013xca} for the $d=10$ case), but  its possible application to integrable deformations had been overlooked. We note that Lunin-Maldacena deformations of $d=11$ supergravity have been, starting with the original \cite{Lunin:2005jy}, studied in a number of papers \cite{Ahn:2005vc, Gauntlett:2005jb, Berman:2007tf, CatalOzer:2009xd,Hellerman:2012zf,Orlando:2018kms}, however, only recently has the tri-vector description become clearer in the context of generalized geometry, more accurately exceptional Sasaki-Einstein structure \cite{Ashmore:2018npi}.  

The point of this paper is that there is a simple transformation that does the job without all this additional structure, and at least for ExFT with SL(5) symmetry group, it is known \cite{Blair:2014zba}. More precisely, there is a frame change and, as we will explain, demanding that the generalized metric is invariant under this operation, we arrive at the required map. The next step is to specify the deforming tri-vector in the ``non-geometric" frame, and following the close analogy to YB deformations in $d=10$ \cite{Bakhmatov:2017joy}, we assume it is an antisymmetric product of Killing vectors, or simply ``tri-Killing": 
\begin{equation} \label{trivector}
\W^{abc} = \frac{1}{3!}\, \r^{\a\b\g} K^a_\a K^b_\b K^c_\g,
\end{equation}
where $\r^{\a\b\g}$ is constant and totally antisymmetric and $K_{\alpha}$ are Killing vectors of the background $d=4$ internal space. Note that assuming a U(1) isometry with Killing vector $\partial_{z}$, we easily recover the bi-vector $\Omega = \partial_{z} \wedge \Theta$, where $\rho$ reduces to the $r$-matrix, $ \rho^{z \a \b} = r^{\a \b}$. At this point, we just have to restrict $\r$ in a bid to ensure a solution to $d=11$ supergravity exists. Following the analogy to $d=10$ further, while working with explicit examples, we will see that the necessary conditions, at least in the SL(5) theory, are the vanishing of $R$-flux and the tracelessness of $Q$-flux. The former is an analogue of the CYBE for rank 3 matrices, while the latter is a $d=11$ version of the unimodularity condition \cite{Borsato:2016ose}, namely the constraint that distinguishes solutions of usual and generalized supergravity. Ensuring both conditions hold, we recover examples in the Lunin-Maldacena class, which as we explain in the appendix \ref{ehlers}, correspond to classic Ehlers-type \cite{Ehlers:1961zza} solution generating transformations in $d=8$. 

The structure of this paper is as follows. Section~\ref{exft} contains the basic review of the necessary facts from the SL(5) ExFT. In the section~\ref{section:def} we derive the deformation prescription and explain the roles played by the $R$ and $Q$-fluxes. Section~\ref{examples} presents some examples, and we conclude in section~\ref{disc}. In appendix \ref{QR} we derive conditions that follow from vanishing of the $R$-flux and trace of the $Q$-flux. In appendix \ref{ehlers} the connection between Lunin-Maldacena (TsT) transformations \cite{Lunin:2005jy} and Ehlers transformations in pure gravity \cite{Ehlers:1961zza} is explained.

\section{Exceptional Field Theory: \texorpdfstring{$\operatorname{SL}(5)$}{SL(5)} group}\label{exft}

In this section, we briefly explain the ExFT with SL(5) symmetry group which will be used 
to illustrate our map. This theory corresponds to the dimensional reduction of $d=11$ supergravity on a four-torus and has a good balance between high enough internal space dimension, and simple enough form of generalized metric. The theory is  formulated on a $(7+10)$-dimensional space-time, split between 7 external directions with Lorentzian signature,\footnote{Using time-like U-dualities one can construct external spaces of Euclidean signature, while the time direction is in the internal space~\cite{Malek:2013sp}, more on this below.} where one recovers gauged supergravity, and 10 extended internal directions parametrized by coordinates $\XX^{mn}$ transforming in the $\bf 10$ of SL(5). The fundamental representation of SL(5) is labelled by Latin indices from the middle of the alphabet, $m,n$, etc.

The coordinates can be divided into normal, ``geometric" coordinates $x^a=\XX^{5a}$ and dual, ``non-geometric" coordinates $y_{ab}=\e_{abcd}\XX^{cd}$, where $a,b,\dots =1,\dots,4$. To keep the total number of space-time dimensions at eleven, and also for consistency of generalized diffeomorphism transformations, one needs to impose a section constraint \cite{Berman:2012vc}. Various solutions of this constraint lead to $d=11$, Type IIA, Type IIB, $d=7$ ungauged or gauged supergravities (see \cite{Musaev:2015ces} and \cite{Hohm:2013jma} for further details). In what follows we will always assume the geometric solution of the section constraint, so all fields depend only on the geometric coordinates $\XX^{5a}$.

The field content of the theory encodes the field content of $d=11$ supergravity in the $7+4$ split and consists of the external metric $g_{\m\n}$, where Greek letters from the middle of the alphabet are used, $\m,\n=1,\ldots, 7$; $10$ vectors $\mA_\m{}^{mn}$ playing the role of a gauge connection, five two-forms $\mB_{\m\n m}$ and the generalized metric $\mM_{mn}$ which is an element of the coset $SL(5)/SO(5)\times \RR^+$. The full action for the theory, invariant under external diffeomorphisms, internal generalized diffeomorphisms, and all gauge transformations has been constructed in \cite{Musaev:2015ces}. However, in what follows we will be working exclusively in the truncated version of the SL(5) ExFT~\cite{Blair:2014zba}, completely ignoring the tensor fields. At the level of supergravity this implies that we are only allowed to consider initial backgrounds without off-diagonal blocks in the metric in the $11=7+4$ split. The equations of motion for the internal space fields $g_{ab}, C_{abc}$ are the usual supergravity ones, and are encoded in the dynamics of the generalized metric $\mM_{mn}$.

Thus, we will focus on the generalized metric, which treats the metric $g_{ab}$ of the four-dimensional internal space and the components of the three-form field $C_{abc}$ on an equal footing.\footnote{Here we use the parametrization of \cite{Blair:2014zba}, an alternative parametrization can be found in \cite{Park:2013gaj} A general approach to deriving generalized metrics from the $E_{11}$ decomposition has been described in \cite{Berman:2011cg}.} It will be convenient to introduce the dual vector on the internal space
\begin{equation}
\label{V}
V^{a}=\fr1{3!}\fr{1}{\sqrt{g}}\e^{abcd}C_{bcd},
\end{equation}
where $\e^{abcd}$ is an absolutely antisymmetric symbol and $\e^{1234}=1$. The generalized metric may be expressed as \cite{Berman:2011jh}
\begin{equation}
\label{eq:LittleGenMetric}
\mM_{mn} = |g_7|^{-1/14}
\begin{bmatrix}
|g|^{-1/2}g_{ab} && V_a \\ \\
V_b && \pm |g|^{1/2}(1 \pm V^2)
\end{bmatrix}.
\end{equation}
Note that $V^2 = g_{ab} V^a V^b$ and the overall factor in the generalized metric represents the determinant of the seven-dimensional external metric, $g_7 = \det g_{\m\n}$,
which is required in order to make $\mM_{mn}$ transform as a generalized tensor~\cite{Blair:2014zba}. Alternatively, this can be seen from the fact that the external metric $g_{\m\n}$ transforms as a scalar with non-zero weight under generalized diffeomorphism \cite{Musaev:2015ces}. We refer to this representation of $\mM_{mn}$ as ``C-frame". 

The upper signs (plus) in the generalized metric above correspond to the case where the internal space metric $g_{ab}$ is of Euclidean signature, while the external space has one time direction. The lower signs (minus) are to be used in the opposite case, which has been worked out in detail in \cite{Malek:2013sp}. In this case the extended space includes a four dimensional space-time of Lorentzian signature, the U-duality group is still SL(5) and the information about the signature is contained in the local duality group. The latter is SO(3,2) and the generalized metric $\mM_{mn}\in {\rm SL(5)} / {\rm SO(3,2)}$.

In principle the only information one has about the generalized metric in ExFT is that it is an element of the coset space ${\rm SL(5)/SO(5)}\times \RR^+$, and hence the matrix \eqref{eq:LittleGenMetric} presents only a particular choice of parametrization of a coset element. The theory is equally consistent for any other chosen parametrization of the coset space, and alternatively we can choose to write it in the $\W$-frame: 
\begin{equation}
\label{eq:NonGeomGenMetric}
\begin{aligned}
\mM_{mn} = |G_7|^{-1/14}
\begin{bmatrix}
|G|^{-1/2} (G_{ab} \pm W_a W_b) && W_a\\ \\
W_b && \pm |G|^{1/2}
\end{bmatrix},
\end{aligned}
\end{equation}
where we have defined
\begin{equation}\label{W}
W_{a} = \frac{1}{3!} \sqrt{G}\, \e_{abcd}\W^{bcd}
\end{equation}
and $G_{ab}$ is the four-dimensional metric in the new $\W$-frame. The plus and minus signs depend on the signature as before, and $G_7 = \det G_{\m\n}$ is the external metric determinant. One is free to substitute the generalized metric of the $\W$-frame into the action of ExFT and obtain a reformulation of (a sector of) $d=11$ supergravity in terms of the new fundamental fields $G_{ab},\W^{abc}$ instead of $g_{ab},C_{abc}$. This is an eleven-dimensional analogue of $\b$-supergravity of Andriot et al. \cite{Andriot:2012wx,Andriot:2012an}, which provides the natural language for the description of non-geometric backgrounds of M-theory. 

Now the key point is that, we can move between frames by simply equating the generalized metric, 
\begin{equation}\label{gen-metric-L}
\mM_{mn} = |g_7|^{-1/14}
\begin{bmatrix}
|g|^{-1/2}g_{ab} && V_a \\ \\
V_b && \pm|g|^{1/2}(1 \pm V^2)
\end{bmatrix} = 
|G_7|^{-1/14}
\begin{bmatrix}
|G|^{-1/2} (G_{ab} \pm W_a W_b) && W_a\\ \\
W_b && \pm |G|^{1/2}
\end{bmatrix}. 
\end{equation}
As we will explain in the following section, this identification serves as the basis of a map that allows to relate the two metrics, three-form potential and tri-vector. This is the $d=11$ analogue of the open-closed string map, but instead of a bi-vector, the deformation will now be specified by a tri-vector. 

In $d=10$, it was noted that non-geometric fluxes, namely $Q$ and $R$-flux played a special role. The vanishing of the trace of the former is related to the unimodularity condition \cite{Borsato:2016ose}, which determines whether the deformed geometry is a solution to usual supergravity, or the generalized supergravity. In $d=10$, the vanishing of $R$-flux reproduces the homogeneous CYBE, so we can expect both of them to play a special role here for deformations in $d=11$. In the context of the SL(5) ExFT, the $Q$ and $R$-flux are encoded in the generalized torsion of the theory, alongside with the four-form flux sourced by the M5 brane, and other geometric and non-geometric quantities. The generalized torsion belongs to the representation $\bf 10+ 15 + 40$ of SL(5), which upon solving the section constraint decomposes into a set of irreps of $\rm GL(4)\in SL(5)$~\cite{Blair:2014zba}. Among them one finds $\bf 6_2+10_2$, corresponding to the $Q$-flux $Q_a{}^{bcd}$, as well as $\bf \bar{4}_7$, which is the $R$-flux $R^{a,bcde}$, where the subscript denotes weight with respect to  the $\RR$ subgroup of $GL(4)$.

In terms of the tri-vector, the $Q$ and $R$-fluxes can be written as  
\begin{subequations}\label{QR-fluxes}
	\begin{align}
	R^{a,bcde}&=\W^{af[b}\nabla_f\W^{cde]},\label{R-flux-def}\\
	Q_{a}{}^{bcd}&=\nabla_a \W^{bcd},\label{Q-flux-def}
	\end{align}
\end{subequations}
where $\nabla_{a}$ is the covariant derivative consistent with the original internal metric $G_{ab}$.  Note that the expressions are given in the $\W$-frame and upon dimensional reduction, they will recover expressions in the $\b$-frame of DFT. In the section~\ref{examples} we will discuss examples allowing us to tease apart the role of $Q$-flux and $R$-flux in the SL(5) theory. 
It is evident that the generalized metrics in different frames are apparently equal to each other as $5\times 5$ matrices, while the fields, for example the four-dimensional metric, are different. The interpretation of the field $\W^{abc}$ as a fundamental field of the eleven-dimensional theory implies that one has the tri-vector field instead of the three-form field $C_{abc}$. The latter couples magnetically to the M5-brane, while the former interacts magnetically with the $5^3$-brane. Viewing $\W^{abc}$ as a fundamental field provides a geometric way of writing down non-geometric backgrounds \cite{Andriot:2012an,Hassler:2013wsa,Bakhmatov:2016kfn}. This is the same picture as one observes in $\b$-supergravity, where the bi-vector field $\b \in \wedge^2 TM$ provides the gauge potential for the $5_2^2$-brane \cite{Chatzistavrakidis:2013jqa}.

An important point in this work is that we adopt an alternative interpretation of the tri-vector field $\W^{abc}$.  Rather than 
being a dynamical fundamental field, we view it as a parameter encoding a deformation of the pure gravity background with the metric $(G_{\m\n},G_{ab})$. This idea is a straightforward generalization of the approach developed in \cite{Bakhmatov:2018bvp}, based on earlier work \cite{Araujo:2017jkb, Araujo:2017jap, Araujo:2017enj}, where the $\b$-field of DFT was understood as a deformation parameter. This can be viewed as a further continuation of the logic of the CYBE/gravity correspondence~\cite{Matsumoto:2014cja,Matsumoto:2014nra}.

\section{Tri-vector deformation}\label{section:def}

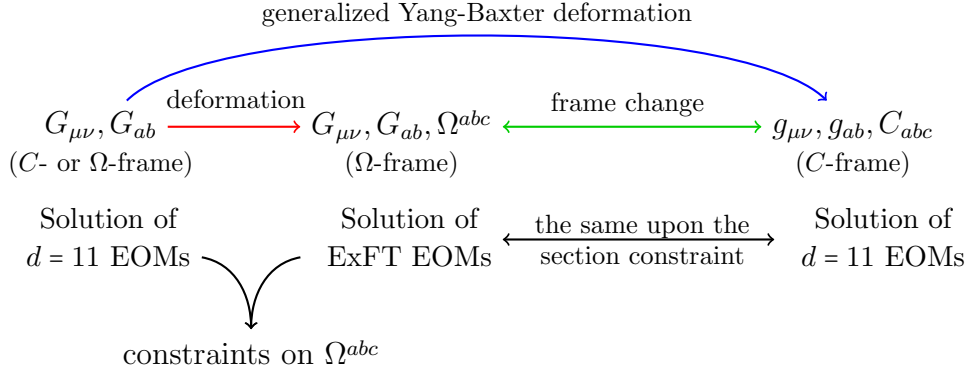
\begin{figure}[ht]
	\centering
	\begin{tikzpicture}
	\node at (0,6) (Gf) {$G_{\m\n},G_{ab}$};
	\node at (4,6) (GQf) {$G_{\m\n},G_{ab}, \W^{abc}$};
	\node at (10,6) (gBf) {$g_{\m\n},g_{ab}, C_{abc}$};
	\node at (0,5.5)  {\footnotesize ($C$- or $\W$-frame)};
	\node at (4,5.5)  {\footnotesize ($\W$-frame)};
	\node at (10,5.5)  {\footnotesize ($C$-frame)};
	\node at (0,4.5) (EOMs0) {\parbox{2.4cm}{\begin{tabular}{c}\small Solution of\\ \small $d=11$ EOMs\end{tabular}}};
	\node at (4,4.5) (bEOMs) {\parbox{2.4cm}{\begin{tabular}{c}\small Solution of\\ \small ExFT EOMs\end{tabular}}};
	\node at (10,4.5) (EOMs) {\parbox{1.8cm}{\begin{tabular}{c}\small Solution of\\ \small $d=11$ EOMs\end{tabular}}};
	\node at (2,3) (cstr) {constraints on $\W^{abc}$};
	
	\node at (1.8,6) [above=0.1cm] {\footnotesize deformation};
	\node at (7,6) [above=0cm] {\footnotesize frame change};
	\node at (5,7.5)  {\footnotesize generalized Yang-Baxter deformation};
	\node at (7.2,4.5) [] {\parbox{3cm}{\centering \footnotesize the same upon the \\ section constraint}};

	\draw[->, thick, red] (Gf) -- (GQf);
	\draw[<->, thick, green!80!black] (GQf) -- (gBf);
	\draw[<->, thick] (EOMs) -- (bEOMs);
	
	\path[->, thick] (EOMs0) edge[out=-10, in=90] (cstr);
	\path[->, thick] (bEOMs) edge[out=190, in=90] (cstr);
	\draw[->, thick, blue] (Gf) .. controls (1.5,7.5) and (8.5,7.5) .. (gBf);
	
	\end{tikzpicture}
	\caption{
		The relationship between a background $(G_{\m\n},G_{ab})$ and its deformation $(g_{\m\n},g_{ab}, C_{abc})$. Starting with a background with $C_{abc}=0$ one switches to the $\W$-frame and simply turns on $\W^{abc}$ (red arrow). Switching to the $C$-frame one obtains a deformed background (green arrow).
	}
	\label{fig:def}
\end{figure}

As explained, we will be exploiting the uniqueness of the SL(5) generalized metric and this will allow us to produce a deformed background from the original data. Upon reduction to $d=10$ and DFT, this rewriting is equivalent to the open-closed string map. For simplicity, we consider original solutions $(G_{\m\n},G_{ab})$ with no three-form on the internal space. After turning on the deformation $\W$ these can be understood as the $\W$-frame backgrounds and transition to the $C$-frame will deform both seven dimensional and internal metrics. Once we have read off the deformed solution $(g_{\m\n},g_{ab},C_{abc})$, we will use the equations of motion of $d=11$ supergravity to confirm the role of $R$-flux and $Q$-flux. Figure~\ref{fig:def} schematically summarizes the idea and reader may take note that the transformation provides simple, closed expressions for the transformed fields.

Let us now unravel the information in \eqref{gen-metric-L}, so that we can directly present the transformation in terms of the fundamental fields: the original metric $(G_{\m\n}, G_{ab})$, split across external and internal spaces, the deforming tri-vector $\W^{abc}$ with legs only along the internal space. Consequently, the deformed metric $(g_{\m\n}, g_{ab})$ is again split and the three-form $C_{abc}$ has legs only along the internal space. This transformation has appeared earlier in~\cite{Blair:2014zba} in the context of the non-geometric flux interpretation for $\W$. 

Equating determinants in both frames we obtain the transformation law for the determinant of the external metric,
\begin{equation}
g_7 = K^{-7/3}\,G_7,
\end{equation}
where we have denoted $g /G = K^{5/3}$. Equivalently, noting that the external metrics in $C$ or $\W$ frames can only differ by a conformal factor, we learn that
\begin{equation}\label{ext}
g_{\m\n} = K^{-1/3} G_{\m\n}.
\end{equation}
One can then eliminate external metrics in~\eqref{gen-metric-L}, which boils down to the relations
\begin{subequations}\label{eqabc}
	\begin{align}
	g_{ab} &= K^{2/3} \left( G_{ab} \pm W_a W_b \right), \label{eqa}\\
	V_a &= K^{-1/6} W_a, \label{eqc}\\
	K &= \left( 1 \pm V^2 \right)^{-1} = \left( 1 \pm W^2 \right)^{-1}. \label{eqb}
	\end{align}
\end{subequations}
In the second equality of \eqref{eqb} we used \eqref{eqa} and 
the algebraic identity
\begin{equation}
\det \left( \d_a^b \pm W_a W^b \right) = 1 \pm W_a W^a.
\end{equation}
Understanding $K$ as a function of the deformation parameter $W_a$, the other two equations in~\eqref{eqabc} express the deformed fields in terms of the original metric $G_{ab}$ and $W_a$. Extracting the three-form and the tri-vector from the dual quantities gives a simple relation:
\begin{equation}\label{Cdef}
C_{abc} = K \W_{abc}.
\end{equation}
Note that $\W_{abc} = G_{aa'} G_{bb'} G_{cc'} \W^{a'b'c'}$.

Now, this has been more or less a mechanical exercise. We still need to specify the tri-vector, and in principle one should require that both $(G_{\m\n},G_{ab})$ and the deformed background $(g_{\m\n},g_{ab},C_{abc})$ are solutions to $d=11$ supergravity. This would produce the most general constraints on the deformation parameter $\W^{abc}$, see Figure~\ref{fig:def}. Leaving that for a separate study, here we will instead use a specific ansatz for the tri-vector, assuming that it is an antisymmetric product of Killing vectors \eqref{trivector}:
\begin{equation}
\W^{abc} = \frac{1}{3!}\, \r^{\a\b\g} K^a_\a K^b_\b K^c_\g.
\end{equation}
Here the constant coefficients $\r^{\a\b\g}$ are, following the analogy to $d=10$, at least required to satisfy the vanishing of $R$-flux~\eqref{R-flux-def}. We derive this condition in appendix \ref{QR}  (\emph{cf}. \eqref{R-flux=0}):
\begin{equation} \label{Rvanishing}
6 \r^{\a\b[\g} \r^{\d\e|\z|} f_{\a\z}{}^{\h]} + \r^{[\g\d\e} \r^{\h]\a\z} f_{\a\z}{}^{\b} = 0, 
\end{equation}
where $f_{\a\b}{}^{\g}$ are the structure constants, defined in the usual way $[K_{\a}, K_{\b}]=f_{\a \b}{}^{\g} K_{\g}$. Greek letters from the beginning of the alphabet are used for the isometry algebra indices.
The vanishing of the trace of $Q$-flux \eqref{Q-flux-def}, upon assuming the tri-Killing ansatz for $\Omega^{abc}$, gives (see appendix  \ref{QR})
\begin{equation}\label{trQ=0}
\r^{\a\b[\g}f_{\a\b}{}^{\d]}=0.
\end{equation}

Following the analogy with $d=10$, one could expect \eqref{Rvanishing} to constrain the tri-vector in such a way, that the deformed background is a solution to supergravity. In this sense this equation may be viewed as a generalization of the CYBE. However, unlike the $d=10$ case, in the context of SL(5) ExFT the $R$-flux condition is a necessary consequence of the $\text{Tr}\,Q = 0$ condition (\emph{cf}.~appendix~\ref{QR}). This comes about because our tri-vector is restricted to a four-dimensional internal space. Moreover, at the level of deformed $d=11$ geometry, the supergravity field equations require both~\eqref{Rvanishing}~\emph{and}~\eqref{trQ=0}. This has been checked on numerous examples, some of which appear in the next section. It is an open question if one can disentangle vanishing $R$-flux and tracelessness of $Q$-flux conditions for larger U-duality groups. We hope to return to this problem in future.

Continuing the $d=10$ analogy, the characteristic vector field $I$ of generalized supergravity is precisely given by $\textrm{Tr}\,Q$. Thus vanishing of the trace of $Q$ restricts the deformation to be a solution to usual (and not generalized) supergravity. It is not known if any meaningful generalized supergravity can be defined in $d=11$, hence it is natural to expect $\textrm{Tr}\,Q$ to vanish for \emph{all} $d=11$ deformations. This ultimately restricts us to examples in the Lunin-Maldacena class. As explained in \cite{Lunin:2005jy}, there is a $d=8$ description underlying the transformations in $d=11$ (see also comments in appendix \ref{ehlers}), which implicitly assumes a reduction on a three-torus. In effect, this means that there is a U(1) Killing direction $\partial_z$ along the M-theory circle.  With the factorization $\Omega = \partial_z \wedge \Theta$ or  $\rho^{z \a \b} = r^{\a \b}$, it is easy to recover the $d=10$ YB deformations. The second term in the $d=11$ $R$-flux condition~\eqref{Rvanishing} vanishes identically using $f_{z\a}{}^\b = 0$, and the first term reproduces the CYBE for $r^{\a\b}$. The $d=11$ $\textrm{Tr}\,Q=0$ condition~\eqref{trQ=0} leads to a nontrivial relation: $r^{\a\b}f_{\alpha\beta}{}^{\gamma} = 0$, where the indices $\alpha,\beta,\gamma$ do not include $z$. That is, vanishing of the Tr $Q$ flux in $d=11$ yields the unimodularity condition \cite{Borsato:2016ose} in $d=10$.

Keeping in mind this discussion of the constraints on the tri-vector, we can summarize the deformation procedure 
in the following steps. 
\begin{itemize}
	\item Start with a solution where the metric is split into $d=7$ and $d=4$ blocks given by $(G_{\m\n}, G_{ab})$ and choose a tri-vector field $\W^{abc}$ of the form \eqref{trivector}; 
	\item The deformed external metric is $g_{\m\n} = K^{-1/3} G_{\m\n}$~\eqref{ext} and the deformed internal metric is $g_{ab} = K^{2/3} \left( G_{ab} \pm W_a W_b \right)$~\eqref{eqa}, where the auxiliary one-form $W = \star_4 \W$~\eqref{W} and $K = \left( 1 \pm W_a W^a \right)^{-1}$; 
	\item Finally, the three-form generated through the deformation is simply $C_{abc} = K \W_{abc}$.
\end{itemize}
Recall that the upper (lower) signs correspond to the Euclidean (Lorentzian) signature of the internal metric. Note that as demonstrated by the $AdS_4 \times S^7$ example of section~\ref{section:ads}, the initial space-time with metric $G_{ab}$ is in fact allowed to be accompanied by a nonzero four-form flux $F$, as long as it has legs in the external space only. In that case the original flux contributes to the deformation alongside $C_{abc} = K \W_{abc}$:
\begin{equation}\label{flux}
\tilde F = \textrm{d} C + \left( 1 - \W \righthalfcup \ast_{11} \right) F.
\end{equation}
The last term is needed for consistency of the deformed solution, and can be understood in the $d=11$ TsT setup~\cite{Lunin:2005jy,CatalOzer:2009xd}. After reducing along $z$ to IIA theory, one performs a $d=10$ TsT deformation w.r.t.\ the remaining two generators of $\W$. As one passes through type IIB stage, a self-dual completion of the RR 5-form must be added for consistency. This gives rise to the last term in~\eqref{flux}.

\section{Examples}\label{examples}

For space-times possessing at least three commuting isometries $K_1, K_2, K_3$, \eqref{Rvanishing} and \eqref{trQ=0} are automatically satisfied by the deformations with a tri-vector $\W \sim K_1 \wedge K_2 \wedge K_3$. These are equivalent to $d=11$ TsT deformations. Below we will demonstrate this in two basic examples of flat space and the Lunin-Maldacena deformation of $AdS_4 \times S^7$. We will conclude this section with an example of a non-abelian deformation. We remind the reader again for reasons presented earlier that all deformations in $d=11$ are restricted to the TsT class.

\subsection{TsT in flat space}

As a preliminary sanity check of our deformation prescription, we consider space-times with a three-torus factor and without four-form flux:
\begin{equation}
\textrm{d} s^2 = \textrm{d} s^2(M_7) + G_{zz} \textrm{d}z^2 + \d_{ij}\, \textrm{d} x^i \textrm{d} x^j,
\end{equation}
where $M_7$ is some Ricci-flat pseudo-Riemannian manifold. The coordinates $x^a = (z, x^i)$ for $i=1,2,3$ parametrize the internal space of the SL(5) ExFT. With an elementary choice of the deformation tri-vector
\begin{equation}
\W = \gamma\, \partial_{x^1} \wedge \partial_{x^2} \wedge \partial_{x^3} \quad \Rightarrow\quad W = \g\, G_{zz}^{1/2} \textrm{d} z,
\end{equation}
the deformed internal metric is given by
\begin{equation}
\begin{split}
\textrm{d}\tilde s_{(4)}^2 
&= K^{2/3} \left(G_{ab}\, \textrm{d} x^a \textrm{d}x^b + W^2\right)  = K^{2/3} \left[ \d_{ij}\, \textrm{d}x^i \textrm{d}x^j + G_{zz} \textrm{d}z^2 + \g^2 G_{zz} \textrm{d}z^2 \right] \\
&= K^{-1/3}G_{zz} \textrm{d}z^2 + K^{2/3} \d_{ij}\, \textrm{d}x^i \textrm{d}x^j,
\end{split}
\end{equation}
where $K = (1 + \gamma^2)^{-1}$ is constant. Recalling the deformation law of the external metric~\eqref{ext}, the complete $d=11$ deformed background can be written as
\begin{equation}\label{deformed}
\textrm{d} \tilde s_{(11)}^2 = K^{-1/3} \left[ \textrm{d} s^2(M_7) + G_{zz} \textrm{d} z^2 \right] + K^{2/3} \d_{ij}\, \textrm{d} x^i \textrm{d} x^j.
\end{equation}
There is no four-form flux after the deformation, because the emergent three-form potential~\eqref{Cdef} is constant, 
\begin{equation}
C = \frac{\g}{1 + \g^2}\, \textrm{d} x^1 \wedge \textrm{d} x^2 \wedge \textrm{d} x^3.
\end{equation}
Thus the deformation is simply a blockwise rescaling of the metric by constants $K^{-1/3},K^{2/3}$, which of course agrees with the one-parameter deformation according to the TsT prescription~\cite{Lunin:2005jy,CatalOzer:2009xd,Deger:2011nb}. Admittedly, this was a completely trivial exercise, so let us move onto a more meaty example.

\subsection{Lunin-Maldacena deformation of \texorpdfstring{$AdS_4 \times S^7$}{AdS4xS7}}\label{section:ads}

Let us perform one more consistency check by recovering the supersymmetry preserving deformation of Lunin and Maldacena \cite{Lunin:2005jy}. This example is non-trivial in the sense that the R-symmetry U(1), generated by  $\partial_{\psi}$, which allows one to preserve $\mathcal{N}=2$ in three dimensions, cannot be touched by the deformation. We are aided by the fact that the corresponding tri-vector has already been identified in \cite{Ashmore:2018npi}, so here we will now show how it arises from our prescription. By choosing different combinations of the U(1)'s, it is possible to find more non-supersymmetric deformations by the same procedure.

To get started, we recall that the original geometry may be expressed as 
\begin{equation}
\textrm{d} s^2 = \frac{1}{4} \textrm{d} s^2 (AdS_4) + R^2 d\W_{(7)}^2, \quad
F_4 = \frac{3}{8R} \text{vol}_{AdS_4},  
\end{equation}
where using the notation $s_\a = \sin\a$, etc.,  we can represent the unit metric on the seven-sphere as
\begin{equation}
\textrm{d} \W_{(7)}^2 = \textrm{d} \th^2 + s_\th^2 (\textrm{d}\a^2 + s_\a^2\, \textrm{d} \b^2) + c_\th^2\, \textrm{d} \f_1^2 + s_\th^2 \left[ c_\a^2\, \textrm{d} \f_2^2 + s_\a^2 \left( c_\b^2\, \textrm{d}\f_3^2 + s_\b^2\, \textrm{d}\f_4^2 \right) \right].
\end{equation}
Now, the four-dimensional submanifold that will play a role of the internal space of the ExFT is simply spanned by the Cartan generators U(1)$^4 \subset$~SO(8):
\begin{equation}
G_{ab} \textrm{d} x^a \textrm{d}x^b = R^2 \left\{ c_\th^2\, \textrm{d}\f_1^2 + s_\th^2 \left[ c_\a^2\, \textrm{d}\f_2^2 + s_\a^2 \left( c_\b^2\, \textrm{d}\f_3^2 + s_\b^2\, \textrm{d}\f_4^2 \right) \right]\right\},\quad
\sqrt{G} = R^4 c_\th s_\th^3 c_\a s_\a^2 c_\b s_\b. 
\end{equation}
However, some care is required to separate the R-symmetry direction $\psi$ from the other directions \cite{Lunin:2005jy}: 
\begin{equation}
\begin{aligned}
\psi &= \frac{1}{4} ( \phi_1 + \phi_2 + \phi_3 + \phi_4), \quad \varphi_1 = \frac{1}{4} ( 3 \phi_4 - \phi_1 - \phi_2 - \phi_3), \\
\varphi_2 &= \frac{1}{2} (\phi_3 + \phi_4 - \phi_1 - \phi_2), \quad \varphi_3 = \frac{1}{4} ( 3 \phi_1 - \phi_2 - \phi_3 - \phi_4).  
\end{aligned}    
\end{equation}
Setting aside the R-symmetry, we are left with the three directions to construct the tri-vector
\begin{equation}
\Omega = - \gamma\, \partial_{\varphi_1} \wedge \partial_{\varphi_2} \wedge \partial_{\varphi_3} = \g \left( \p_{\f_1} \wedge \p_{\f_2} \wedge \p_{\f_3} - \p_{\f_1} \wedge \p_{\f_2} \wedge \p_{\f_4} + \p_{\f_1} \wedge \p_{\f_3} \wedge \p_{\f_4} - \p_{\f_2} \wedge \p_{\f_3} \wedge \p_{\f_4}\right),
\end{equation}
which coincides with the tri-vector reported in \cite{Ashmore:2018npi}. At this stage we can be confident that everything is going to work out and, up to coordinates, we are guaranteed to recover the Lunin-Maldacena geometry with $\mathcal{N}=2$ supersymmetry. Nevertheless, let us proceed with the details.

The dual one-form $W = -4 \gamma \sqrt{G}\, \textrm{d}\psi$ gives the deformation factor:
\begin{equation}
K^{-1} = 1 + G^{ab} W_a W_b =  1 + \g^2 R^6 s_\th^6 s_\a^2 \left[ c_\th^2 c_\a^2 + s_\a^2 c_\b^2 s_\b^2 \left( c_\th^2 + s_\th^2 c_\a^2 \right) \right], 
\end{equation}
which again agrees with \cite{Lunin:2005jy}. 
With these values the prescription for the deformed metric leads to:
\begin{equation}
\begin{split}
\textrm{d} \tilde s_{(11)}^2 &= \frac14 K^{-1/3} \textrm{d} s^2(AdS_4) + R^2 K^{-1/3} \left[  \textrm{d} \th^2 + s_\th^2 (\textrm{d} \a^2 + s_\a^2\, \textrm{d} \b^2) \right] \\&+ R^2 K^{2/3} \left\{ c_\th^2\, \textrm{d} \f_1^2 + s_\th^2 \left[ c_\a^2\, \textrm{d}\f_2^2 + s_\a^2 \left( c_\b^2\, \textrm{d} \f_3^2 + s_\b^2\, \textrm{d}\f_4^2 \right) \right] + 16\g^2 G \,\textrm{d} \psi^2 \right\}.
\end{split}
\end{equation}
In order to construct the deformed four-form flux we compute
\begin{equation}
\begin{split}
C = \g R^6 K &\left[ (\textrm{d} \f_1 \wedge \textrm{d}\f_2 \wedge \textrm{d} \f_3) \,c_\th^2 s_\th^4 c_\a^2 s_\a^2 c_\b^2 - (\textrm{d}\f_1 \wedge \textrm{d}\f_2 \wedge \textrm{d}\f_4) \,c_\th^2 s_\th^4 c_\a^2 s_\a^2 s_\b^2 \right.\\
&\left. +(\textrm{d}\f_1 \wedge \textrm{d}\f_3 \wedge \textrm{d}\f_4) \,c_\th^2 s_\th^4 s_\a^4 c_\b^2 s_\b^2 - (\textrm{d}\f_2 \wedge \textrm{d}\f_3 \wedge \textrm{d}\f_4) \,s_\th^6 c_\a^2 s_\a^4 c_\b^2 s_\b^2 \right].
\end{split}
\end{equation}
The complete four-form after the deformation also receives contributions from the original flux~\eqref{flux}
\begin{equation}
\tilde F = \textrm{d} C + \frac{3}{8R} \text{vol}_{AdS_4} + 24 \g R^6 s_\th^5 c_\th s_\a^3 c_\a s_\b c_\b \left(\textrm{d} \th \wedge \textrm{d} \a \wedge \textrm{d}\b \wedge \textrm{d} \psi\right).
\end{equation}
Up to a choice of notation, this is the Lunin-Maldacena deformation. It is worth stressing again that it can be easily generalized to more elaborate deformations, where no supersymmetry is preserved.

\subsection{A non-abelian deformation}
\label{5}

Finally, let us consider an example of a non-abelian deformation with a single U(1), which in our set-up implies there is a $d=10$ description. As explained earlier, the corresponding tri-vector can be put in the form $\W = \p_z \wedge \Th$, where the bi-vector $\Theta$ is non-abelian (it corresponds to a non-abelian $r$-matrix), whereas $z$ is some coordinate, for all extensive purposes the M-theory circle, such that the shifts $\partial_z$ commute with any of the generators of $\Th$. As discussed (\emph{cf.} discussions below \eqref{trQ=0}) the $d=11$ $\textrm{Tr}\,Q=0$ yields vanishing of trace of $Q$-flux in $d=10$ and vanishing of $R$-flux in $d=11$ yields CYBE for the $d=10$ deformations.

This time we will consider a deformation in the Lorentzian subspace of the $d=11$ background. This will be required to obtain non-trivial solutions after dimensional reduction along $z$, and will serve as a demonstration of the workings of the prescription in the Lorentzian case. We start with a solution to $d=11$ supergravity of a form
\begin{equation}
\textrm{d} s_{(11)}^2 = \h_{ab} \textrm{d} x^a \textrm{d} x^b + \textrm{d} s^2 (M_7),
\end{equation}
where $M_7$ is some Ricci-flat Riemannian manifold and $x^a = (x^i, z)$ for $i=0,1,2$ parametrize the internal space of the SL(5) ExFT. Using the Poincar\'e generators of $(1+2)$-dimensional Minkowski subspace, consider $\W = \Th \wedge \p_z$ with the following $\Th$, parametrized by a constant vector $\t^i = (\a,\b,\g)$:
\begin{equation}
\begin{aligned}
\Theta &= \t^i M_{ij} \wedge P^j \\ 
&= \a \left( M_{01} \wedge P_1 + M_{02} \wedge P_2 \right) + \b \left( M_{01} \wedge P_0 + M_{12} \wedge P_2 \right) + \g \left( M_{02} \wedge P_0 - M_{12} \wedge P_1 \right)\\
&= (\b x^0 - \a x^1)\, \p_0 \wedge \p_1 + (\g x^0 - \a x^2) \,\p_0 \wedge \p_2 + (\g x^1 - \b x^2) \,\p_1 \wedge \p_2.
\end{aligned}
\end{equation}
The CYBE for this $\Theta$ indicates that a valid deformation in $d=10$ exists whenever $\t$ is null: $\a^2 - \b^2 - \g^2 = 0$. The fact that a corresponding solution exists was explicitly checked  in \cite{Araujo:2018rbc}. Using the auxiliary functions
\begin{equation}
\begin{aligned}
W &= (\g x^1 - \b x^2) \textrm{d} x^0 - (\g x^0 - \a x^2) \textrm{d} x^1 + (\b x^0 - \a x^1) \textrm{d} x^2,\\
K &= \left[ 1+ (\g x^1 - \b x^2)^2 - (\g x^0 - \a x^2)^2 - (\b x^0 - \a x^1)^2 \right]^{-1},
\end{aligned}
\end{equation}
the deformed background is given by:
\begin{equation}
\begin{gathered}
\textrm{d} \tilde s_{(11)}^2 = K^{2/3} \left[ \h_{ij}\, \textrm{d} x^i \textrm{d} x^j + \textrm{d} z^2 - W^2 \right] + K^{-1/3} \textrm{d} s^2 (M_7),\\
C = K \left[-(\b x^0 - \a x^1) \,\textrm{d} x^0 \wedge \textrm{d} x^1 - (\g x^0 - \a x^2) \,\textrm{d} x^0 \wedge \textrm{d} x^2 + (\g x^1 - \b x^2) \,\textrm{d} x^1 \wedge \textrm{d} x^2 \right] \wedge \textrm{d} z.
\end{gathered}
\end{equation}
Checking the Einstein equations demonstrates that this background is not a solution unless $\a=\b=\g=0$. This constraint, which trivializes the deformation, is in fact the $\textrm{Tr}\,Q=0$ condition~\eqref{trQ=0}.

We have deliberately picked a tri-vector containing a bi-vector that we know corresponds to a valid deformation in $d=10$ and the resulting geometry is a solution to generalized supergravity. As a result, $Q$-flux in $d=10$ has to pick up a trace to support the solution. This example is essentially testing whether there is the same freedom in $d=11$ and we learn that there is not, since $\a=\b=\g=0$ is telling us that $\textrm{Tr}\,Q=0$ and this kills the deformation.

That being said, one can naively\footnote{The dimensional reduction is ``naive" in the sense that upstairs in $d=11$, there is no solution, but using the usual reduction ansatz, one recovers field content that can be completed to a solution of generalized supergravity.} dimensionally reduce the corresponding geometry, which will give a solution in generalized $d=10$ supergravity. Using the standard reduction ansatz~\eqref{reduction}, we arrive at a $d=10$ background
\begin{equation}
\begin{gathered}
\textrm{d}\tilde s_{(10)}^2 = K \left[ \h_{ij}\, \textrm{d}x^i \textrm{d}x^j -W^2 \right] + \textrm{d}s^2 (M_7), \qquad e^{2\phi} = K,\\
b =  K \left[-(\b x^0 - \a x^1) \,\textrm{d}x^0 \wedge \textrm{d}x^1 - (\g x^0 - \a x^2) \,\textrm{d}x^0 \wedge \textrm{d}x^2 + (\g x^1 - \b x^2) \,\textrm{d}x^1 \wedge \textrm{d}x^2 \right].
\end{gathered}
\end{equation}
This is a solution to generalized supergravity with the vector field
\begin{equation}
I = \nabla \cdot \Th = 2(\a \p_0 + \b \p_1 + \g \p_2),
\end{equation}
as soon as the YB equation for $\Th$ holds, $\a^2 - \b^2 - \g^2 = 0$. We observe that this condition coincides with the $R$-flux vanishing constraint~\eqref{Rvanishing}, whereas the $\textrm{Tr}\,Q=0$ constraint implies that $\a = \b = \g = 0$. Although the deformed background is not a solution to $d=11$ supergravity, it produces a family of solutions to $d=10$ generalized supergravity after dimensional reduction.

\section{Discussion}\label{disc}

In this paper we have developed a prescription for deformations of $d=11$ supergravity, which is a spin-off of the existing literature on SL(5) ExFT. This demonstrates that ExFT correctly reproduces the known TsT deformations in $d=11$. Along the way we have derived algebraic constraints for the rank three deformation tensor,~\eqref{Rvanishing}~and~\eqref{trQ=0}, which presumably are related to higher dimensional analogues of the CYBE.

The main limitation of our work is that so far we have only studied deformations that involve a U(1) direction. In any space-time with a U(1)$^3$ isometry, for a deformation determined by three commuting Killing vectors one can go to adapted coordinates where these vectors generate shifts, and the whole transformation is then always a TsT. This is in line with the YB deformations in $d=10$, where abelian $r$-matrices are equivalent to the usual TsT. Thus in the U(1)$^3$ case our work provides a simple and concise way of performing TsT deformations in $d=11$, or in $d=10$ after reducing along one of the isometry directions. One has to use a simple prescription~\eqref{eqabc}, which hinges on knowing a single piece of data, namely a one-form $W$ constructed from the Killing vectors of the original solution.

When there is only one commuting isometry, the deformation tri-vector has the form $\W = \p_z \wedge \Th$, and we are able to incorporate the solutions to the $d=10$ generalized supergravity into this prescription, as shown by the last example. We saw that the $d=11$ background obtained in the section \ref{5} did not satisfy the supergravity field equations. This is in line with the bi-vector~$\Th$ being non-abelian, and as such not leading to a TsT deformation in $d=10$. Nevertheless, by dimensional reduction one arrives at the YB deformed solution in $d=10$, with the $r$-matrix proportional to $\Theta$. The algebraic condition of $R=0$ that comes from the SL(5) ExFT was shown to reduce to the CYBE for $\Th$.

Such deformations, which in general have non-zero trace of the $Q$-flux, in 10 dimensions fall into the class of generalized supergravity backgrounds, characterized by an extra vector field $I = \nabla \cdot \Theta$. This theory can be embedded into the ExFT~\cite{Sakatani:2016fvh, Baguet:2016prz, Sakamoto:2017wor}. There is no known analogue of generalized supergravity in $d=11$ so far, and the analysis here may give a hint how to construct it, if possible.

The last example also demonstrates that the $\textrm{Tr}\,Q=0$ condition implies the $R=0$ condition, which agrees with the proof given in the appendix~\ref{QR}. This relationship between $\textrm{Tr}\,Q$ and $R$-flux conditions has been observed in an assortment of other sample deformations that we have checked. It has always been the case that the $Q$-traceless condition either coincides with, or provides a solution to the $R=0$ condition.

A natural question arises: whether any non-trivial deformations exist that are intrinsically eleven-dimensional? In other words, solutions that are not merely a TsT or an uplift of the non-abelian YB deformations from $d=10$. In answering this we are restricted by the two main assumptions used in this article: simple isometry algebras of a sphere or a flat space, and the SL(5) ExFT. The latter has four-dimensional internal space, and hence is only capable of accommodating the isometries that are acting within a four-dimensional submanifold. This excludes e.g.\ conformal algebras in dimension higher than four, and we are left with the symmetry groups of a sphere, flat space, and $AdS_4$.

The only tri-vector possible within the Poincar\'e algebra that is completely non-abelian is cubic in momentum generators. For instance one could consider
\begin{equation}
\W = \a (M_{12} \wedge M_{13} \wedge M_{14} + M_{13} \wedge M_{23} \wedge M_{34}) + \b (M_{12} \wedge M_{13} \wedge M_{34} + M_{14} \wedge M_{13} \wedge M_{23}),
\end{equation}
where none of the generators appearing in any of the terms commutes with the other two. The specific directions here are chosen so as to make the $R$-flux condition as simple as possible: $\a^2 = \b^2$ (under the assumption that $x^1$ is timelike). However, the corresponding $d=11$ deformed geometry is not a solution, unless $\textrm{Tr}\,Q = 0$, which imposes $\a=0=\b$. We have observed a similar situation in the non-abelian deformations using the generators of the conformal algebra of $AdS_4$ and suppose that it may be generic.

One possible way out might be to study other ExFT's. For instance, the $E_{7}$ theory has a seven-dimensional internal space, which is large enough to study deformations based on the conformal algebra of any $AdS$ solution in $d=11$ supergravity~\cite{Wulff:2016vqy}. The algebraic argument that $\textrm{Tr}\,Q=0$ implies $R=0$ fails in higher dimensions (\emph{cf}.\ appendix~\ref{QR}), which hints that perhaps one will have more freedom in the deformed geometries as well. We leave this to future work.

It is still an open issue to check if a ``generalized $d=11$ supergravity'' is possible, where $\textrm{Tr}\,Q=0$ is relaxed. Kappa-symmetry of the superstring permits a generalization of $d=10$ supergravity, however it seems that the kappa-symmetry of the supermembrane \cite{Bergshoeff:1987cm} does not allow for an extension of $d=11$ supergravity to a similar generalized $d=11$ supergravity \cite{Wulff:2016tju}. It is interesting to see if examples can be constructed with $\textrm{Tr}(Q) \neq 0$. We saw that this was not possible within the SL(5) theory, but it may be possible for larger U-duality groups. As reviewed in the introduction, in the string theory case the YB deformation narrative admits at least three complementary descriptions: in terms of integrability of the deformed $\sigma$-model, generation of non-trivial solutions to (generalized) supergravity using a simple construction based on solutions to the CYBE, and the non-commutativity of the dual super-Yang-Mills theory in the $AdS_5\times S^5$ examples. In the non-commutative description, the CYBE was shown to be related to the Jacobi identity of the non-commutative algebra associated with the open string endpoint position operators \cite{Araujo:2017jkb, Araujo:2017jap, Araujo:2017enj}. One can speculate that in the $d=11$ case open strings can be replaced with open M2-branes. In that case it is expected that instead of the usual commutators one will have to deal with the three-algebras \cite{Bagger:2012jb, deAzcarraga:2010mr}, where the Jacobi identity is replaced with the Fundamental (or Filippov) Identity (FI) \cite{Filippov}. If, inspired by the $d=10$ string theory case, we identify $\Omega^{abc}(X)=[X^a,X^b,X^c]$, where the bracket is a 3-bracket, one can check that the FI does not match precisely the $R=0$ condition in the SL(5) theory. Nonetheless, one may show that if we contract four indices of the FI with totally antisymmetric symbol $\epsilon^{bcde}$ the FI reduces to $R=0$. Exploring the physical meaning of the generalized CYBE equation~\eqref{Rvanishing} and the 3-algebras is an interesting problem that deserves further investigation.

\section*{Acknowledgements}

We would like to thank A.~\c{C}atal-{\"O}zer, A.~Tseytlin, and L.~Wulff for related discussion.
IB thanks the Istanbul Centre for Mathematical Sciences for hospitality. The work of EtM was funded by the Russian state grant Goszadanie 3.9904.2017/8.9 and by the Foundation for the Advancement of Theoretical Physics and Mathematics ``BASIS''. The work of IB and E\'OC was supported in part by the Korea Ministry of Science, ICT \& Future Planning, Gyeongsangbuk-do and Pohang City. IB and EtM are partially supported by the Russian Government program of competitive growth of Kazan Federal University. MMShJ would like to thank the hospitality of ICTP HECAP where this work was finished and acknowledges the support by 
INSF grant No 950124 and Saramadan grant No. ISEF/M/97219.
NSD wishes to thank IPM for hospitality and gratefully acknowledges financial support of ICTP network scheme NT-04 for this visit.

\appendix
\section{R- and Q-flux}\label{QR}

In this section we derive the algebraic equations~\eqref{Rvanishing} and~\eqref{trQ=0} from the vanishing $R$-flux condition and tracelessness of the $Q$-flux. 

Note that in order to obtain a solution after the deformation one must impose the dynamical equations for $\W^{abc}$ that follow from the ExFT. In case of $\beta$-supergravity the equations of motion were shown to follow from vanishing of $R$-flux and tracelessness of $Q$-flux~\cite{Bakhmatov:2018bvp}. Moreover, these two were independent equations, which reflects the freedom to replace usual supergravity with generalized supergravity. As we will demonstrate now, in the SL(5) ExFT they are not independent. It is still an open question whether the equations of motion of a generic ExFT require both tracelessness of $Q$ and vanishing of $R$-flux.

We begin with the condition $R^{a,bcde}=0$. Greek letters from the beginning of the alphabet are used for the isometry algebra indices. Substituting the tri-Killing ansatz (\ref{trivector})
into the R-flux (\ref{R-flux-def})
one obtains
\begin{equation}
\label{3cybe.1}
\begin{aligned}
R^{a,bcde} &= 3\r^{\a\b\g}\r^{\d\e\z}K_{\a}{}^{a} K_{\b}{}^{g} K_{\g}{}^{[b}K_{\d}{}^{c} K_{\e}{}^{d} \nabla_g K_{\z}{}^{e]}\equiv
3\r^{\a\b\g}\r^{\d\e\z}K_{\a\b\g\d\e}^{g a[bcd}\nabla_g K_{\z}{}^{e]}\\
&= \fr34\r^{\a\b\g}\r^{\d\e\z}\Big[K_{\a\b\g\d\e}^{ga[bcd]}\nabla_g K_{\z}{}^{e}-K_{\a\b\g\d\e}^{ga[ebc]}\nabla_g K_{\z}{}^{d}+K_{\a\b\g\d\e}^{ga[deb]}\nabla_g K_{\z}{}^{c}-K_{\a\b\g\d\e}^{ga[cde]}\nabla_g K_{\z}{}^{b}\Big],
\end{aligned}
\end{equation}
where the antisymmetrization in the four indices $[bcde]$ was decomposed in four terms in the second line, and for the sake of brevity we introduced the notation
\begin{equation}
K_{\a\b\g\d\e}^{abcde}\equiv K_{\a}{}^{a}K_{\b}{}^{b}K_{\g}{}^{c}K_{\d}{}^{d}K_{\e}{}^{e}.
\end{equation}
In order to obtain the structure constants $f_{\a\b}{}^\g$ of the Killing vector algebra, antisymmetrization in $[\a\z]$ must be organized. For the product $\r^{\a\b\g}\r^{\d\e\z}$ this is equivalent to antisymmetrization of the two pair of indices $\b\g$ and $\d\e$. Let us organize this symmetry in the expression \eqref{3cybe.1} and with that goal in mind we start with the following identical transformations of the first term in brackets above
\begin{equation}
\begin{aligned}
&\fr34\r^{\a\b\g}\r^{\d\e\z}\Big[K_{\a\b\g\d\e}^{ga[bcd]}-K_{\a\b\g\d\e}^{g[abcd]}\Big]\nabla_g K_\z{}^{e}+\fr34\r^{\a\b\g}\r^{\d\e\z}K_{\a\b\g\d\e}^{g[abcd]}\nabla_g K_\z{}^{e} \\
=&\ \fr18\r^{\a\b\g}\r^{\d\e\z}\Big[\big(K_{\a\b\g\d\e}^{gabcd} -K_{\a\b\g\d\e}^{gcdab}\big)+\big(K_{\a\b\g\d\e}^{gadbc} -K_{\a\b\g\d\e}^{gbcad}\big)+\big(K_{\a\b\g\d\e}^{gacdb} -K_{\a\b\g\d\e}^{gdbac}\big)\Big]\nabla_g K_\z{}^{e}\\
&+\fr34\r^{\a\b\g}\r^{\d\e\z}K_{\a\b\g\d\e}^{g[abcd]}\nabla_g K_\z{}^{e} \\
=&\ \fr18\Big(\r^{\a\b\g}\r^{\d\e\z}-\r^{\a\d\e}\r^{\b\g\z}\Big)\Big[K_{\a\b\g\d\e}^{gabcd}+K_{\a\b\g\d\e}^{gadbc}+K_{\a\b\g\d\e}^{gacdb}\Big]\nabla_g K_\z{}^{e}+\fr34\r^{\a\b\g}\r^{\d\e\z}K_{\a\b\g\d\e}^{g[abcd]}\nabla_g K_\z{}^{e} \\
=&\ \fr38\r^{\a\b\g}\r^{\d\e\z}K_{\b\h\g\d\e}^{ae[bcd]}f_{\a\z}{}^\h+\fr34\r^{\a\b\g}\r^{\d\e\z}K_{\a\b\g\d\e}^{g[abcd]}\nabla_g K_\z{}^{e} .
\end{aligned}
\end{equation} 
Upon insertion back to \eqref{3cybe.1}, the first term above gains antisymmetrization in $[ebcd]$, which is equivalent to antisymmetrization in $[\h\g\d\e]$. The second term in the last line above is more subtle and can be rewritten as follows
\begin{equation}
\begin{aligned}
\fr34\r^{\a\b\g}\r^{\d\e\z}K_{\a\b\g\d\e}^{g[abcd]}\nabla_g K_\z{}^{e}=&\fr54\r^{\a\b\g}\r^{\d\e\z}K_{\a\b\g\d\e}^{[gabcd]}\nabla_g K_\z{}^{e}-\fr24\r^{\a\b\g}\r^{\d\e\z}K_{\a\b\g\d\e}^{[abcd]g}\nabla_g K_\z{}^{e}\\
=&-\fr14\r^{\a\b\g}\r^{\d\e\z}K_{\a\b\g\d\h}^{[abcd]e}f_{\e\z}{}^\h.
\end{aligned}
\end{equation}
The term with $[gabcd]$ vanishes because the internal indices $a,b,\dots$ for SL(5) theory run from 1 to 4 only. Hence, altogether for the R-flux in \eqref{3cybe.1} we obtain
\begin{equation}
\label{3cybe.2}
R^{a,bcde} = \fr32\r^{\a\b[\g}\r^{\d\e|\z|}f_{\a\z}{}^{\h]}K_{\b\h\g\d\e}^{aebcd} - \r^{[\a\b\g}\r^{\d]\e\z}f_{\e\z}{}^\h K_{\a\b\g\d\h}^{a[bcde]}
\end{equation}
Note that the two antisymmetrizations in the last term above can be reduced to a single antisymmetrization in the lower indices of $K$ plus an additional term:
\begin{equation}
-\r^{[\a\b\g}\r^{\d]\e\z}f_{\e\z}{}^\h K_{\a\b\g\d\h}^{a[bcde]}=-\fr54\r^{\a\b\g}\r^{\d\e\z}f_{\e\z}{}^{\h}K_{\a\b\g\d\h}^{[abcde]}+\fr14\r^{[\a\b\g}\r^{\d]\e\z}f_{\e\z}{}^\h K_{\a\b\g\d\h}^{bcdea}.
\end{equation}
The first term is again zero for SL(5) theory and after relabelling the indices we obtain the final form for the R-flux:
\begin{equation}
R^{a,bcde} = \frac14 \left( 6 \r^{\a\b[\g} \r^{\d\e|\z|} f_{\a\z}{}^{\h]} + \r^{[\g\d\e} \r^{\h]\a\z} f_{\a\z}{}^{\b} \right) K_{\b\g\d\e\h}^{abcde}.
\end{equation}
Hence, R-flux vanishes if 
\begin{equation}\label{R-flux=0}
6 \r^{\a\b[\g} \r^{\d\e|\z|} f_{\a\z}{}^{\h]} + \r^{[\g\d\e} \r^{\h]\a\z} f_{\a\z}{}^{\b} = 0.
\end{equation}

The condition for the $Q$-flux to be traceless naturally comes from counting of non-vanishing components of mixed symmetry potentials interacting with exotic branes \cite{Kleinschmidt:2011vu} and hence can be imposed independently. Using the $Q$-flux definition~\eqref{Q-flux-def}, we have:
\begin{equation}
Q_{a}{}^{abc} = \frac{2}{3!} \r^{\a\b\g}K_\a{}^a K_\b{}^{[b} \nabla_a K_\g{}^{c]}
= \frac{1}{3!} \r^{\a\b\g}f_{\a\b}{}^\d K_\d{}^{[b} K_\g{}^{c]}, 
\end{equation}
where we have used the the fact that $K$ is Killing, i.e.\  $\nabla_a K_\g{}^a=0$. Vanishing of $Q_{a}{}^{abc}$ thus implies
\begin{equation}\label{tr-Q=0}
\r^{\a\b[\g}f_{\a\b}{}^{\d]} = 0.
\end{equation}

Finally, observe a subtle point specific to the SL(5) and O(3,3) theories, specifically, $Q_{a}{}^{abc}\equiv 0$ implies $R^{a,bcde}\equiv 0$. Indeed, substitute $\W^{abc}=-\e^{abcd}W_d$ into the trace part of the Q-flux
\begin{equation}
Q_{a}{}^{abc}=\nabla_a\W^{abc}=-\e^{abcd}\nabla_{a}W_d \stackrel{!}{=}0.
\end{equation}
For the $R$-flux the same gives
\begin{equation}
\e_{bcde}R^{a,bcde}=-3!\e^{abcd}W_b\nabla_c W_d,
\end{equation}
which is obviously zero upon imposing $Q_{a}{}^{abc}=0$. For fluxes of the O(3,3) theory one ends up with the same conclusions. The reason for such behaviour is that the dimensions of the internal space is too small. 

\section{Understanding TsT as an Ehlers transformation} \label{ehlers}
In this appendix we explain the connection between Lunin-Maldacena (TsT) transformations \cite{Lunin:2005jy} and Ehlers transformations in pure gravity \cite{Ehlers:1961zza}, namely both are described by SL$(2, \mathbb{R})$ transformations in a lower-dimensional effective description. This bypasses any need to explain T-duality. Finally, we will relate this to the transformations of $d=11$ supergravity that are the focus of this paper. 

We start by recalling Ehlers transformations in pure gravity. Given a vacuum solution to the Einstein equations, i. e. $R_{\mu \nu} = 0$, with an isometry direction, Ehlers transformations generate new solutions from the old. Let us assume there is a Killing vector $\partial_{t}$ with the metric 
\begin{equation}
\textrm{d} s^2 = - V ( \textrm{d} t + A) \otimes ( \textrm{d} t + A) + V^{-1} \gamma_{ij} \textrm{d} x^i \otimes \textrm{d} x^j, 
\end{equation}
where $V(x^i)$ is a scalar and $A \equiv A_i \textrm{d} x^i $ is a one-form connection. The scalar factors have been specially chosen so that we arrive in Einstein frame after dimensional reduction. We have adopted the time direction for the isometry, but the arguments do not depend on the signature or the dimensionality. However, for concreteness we will assume the space-time is four-dimensional. 

From the $R_{t i} = 0$ component of the Einstein equation, we get 
\begin{equation}
\textrm{d} ( V^2 *_3 F ) = 0, 
\end{equation}
where $F = \textrm{d} A$ and Hodge duality is taken with respect to the metric $\gamma_{ij}$. Locally, we can now define an additional scalar, 
\begin{equation}
V^2 * F = \textrm{d} \omega.  
\end{equation}

Replacing $F$ with its Hodge dual throughout, one arrives at the three-dimensional Lagrangian: 
\begin{equation}
\mathcal{L} = \sqrt{\gamma} \left( R - \frac{1}{2}  \frac{ \partial_i V \partial^i V + \partial_i \omega \partial^i \omega }{V^2} \right). 
\end{equation}
Our target space of the sigma-model is a hyperbolic space, so any transformation of the scalars $(V, \omega)$ that leaves the Lagrangian invariant is a symmetry of the equations of motion.  Therefore, we can redefine 
\begin{equation}
\tau = \omega + i V, 
\end{equation}
so that the metric on the $H^2$ becomes
\begin{equation}
\textrm{d} s^2 = \frac{ \textrm{d} \tau \otimes \textrm{d} \bar{\tau}}{\textrm{Im}(\tau)^2}. 
\end{equation}
The metric is invariant under the SL$(2, \mathbb{R})$ transformation
\begin{equation}
\tau \rightarrow \tau' = \frac{a \tau + b}{c \tau + d}, \quad a d - b c = 1, \quad a, b, c, d \in \mathbb{R}.  
\end{equation}
Ostensibly, there are three parameters, but only one of them is physical as rescalings of the Killing vector and gauge transformations of the scalar potential remove two of these. 

Now, let us turn our attention to TsT \cite{Lunin:2005jy}, which we will discuss initially in $d=10$. The transformation assumes $U(1)^2$ symmetry, so we have a torus. Neglecting one-forms, which will play no role, one starts by entertaining space-times of the form,
\begin{equation}
\label{$d=10$}
\begin{aligned}
\textrm{d} s^2_{10}  &= e^{\frac{1}{6} ( 2 \Phi - C_1 - C_2)} \gamma_{\mu \nu} \textrm{d} x^{\mu} \otimes \textrm{d} x^{\nu} + e^{2 C_1} \textrm{d} \varphi_1^2 + e^{2 C_2} \textrm{d} \varphi_2^2, \\
B &= h \textrm{d} \varphi_1 \wedge \textrm{d} \varphi_2, 
\end{aligned}
\end{equation}
where the scalars $C_i, h$ and $\Phi$ are assumed just to depend on the lower-dimensional coordinates $x^{\mu}$. We will see that it reduces to an SL$(2, \mathbb{R})$ transformation as in the Ehlers case. Again, one dimensionally reduces, but this time on a torus. The lower-dimensional effective Lagrangian becomes: 
\begin{equation}
\label{8d}
\begin{aligned}
\mathcal{L} &= \sqrt{\gamma} \biggl( R - \frac{1}{6} \partial_{\mu} (2 \Phi - C_1 - C_2) \partial^{\mu} ( 2 \Phi - C_1 - C_2) - \frac{1}{2} \partial_{\mu} (C_1 - C_2) \partial^{\mu} (C_1 - C_2) \\
&- \frac{1}{2} \partial_{\mu} (C_1 + C_2) \partial^{\mu} (C_1 + C_2) - \frac{1}{2} \partial_{\mu} h \partial^{\mu} h e^{-2 (C_1+C_2)} \biggr). 
\end{aligned}
\end{equation}
We can now introduce a complex coordinate
\begin{equation}
\tau = h + i e^{C_1 + C_2},   
\end{equation}
observing that $h$ is simply the two-form on the torus and $e^{C_1 + C_2}$ is the volume of the torus. The TsT transformation is an SL$(2, \mathbb{R})$ transformation of the form, 
\begin{equation}
\tau \rightarrow \tau' = \frac{\tau}{ \gamma \tau + 1}, 
\end{equation}
where $\gamma$ is a constant. It is worth taking note that the combinations $2 \Phi - C_1 - C_2$ and $C_1 -C_2$ do not play any role in the transformation. As a result we see that the combination $e^{-2 \Phi} \sqrt{-g}$ is invariant, in line with our expectations for a transformation based on T-duality.  

Now, it is easy to make the connection to the deformations of $d=11$ supergravity discussed in this paper. Assuming the following ansatz upstairs for the metric and three-form, 
\begin{equation}\label{reduction}
\textrm{d} s^2_{11} = e^{-\frac{2}{3} \Phi} \textrm{d} s^2_{10} + e^{\frac{4}{3} \Phi} \textrm{d} z^2, \quad C  = B \wedge \textrm{d} z 
\end{equation}
where the ten-dimensional metric and field $B$ are defined in \eqref{$d=10$}, one can dimensionally reduce on a three-torus to the same eight-dimensional theory \eqref{8d}. Thus, the transformations in this paper are also SL$(2, \mathbb{R})$ transformations that are close cousins of Ehlers transformations.

	\setstretch{1.0}
	
	\bibliographystyle{utphys}
	\bibliography{biblio.bib}
	
\end{document}

%% file: Defs.tex
\def\a{\alpha} 
\def\b{\beta} 
\def\g{\gamma} 
\def\d{\delta} 
\def\e{\epsilon}
 
\def\z{\zeta} 
\def\h{\eta} 
\def\th{\theta}

\def\m{\mu}
\def\n{\nu} 
 
\def\r{\rho}

\def\t{\tau}  
\def\f{\phi}

\def\Th{\Theta}

\def\W{\Omega}

\def\p{\partial}

\def\fr{\frac}

\def\mA{\mathcal{A}}
\def\mB{\mathcal{B}}

\def\mM{\mathcal{M}}

\def\XX{\mathbb{X}}
\def\RR{\mathbb{R}}